\documentclass[letter, 12pt]{article}

\usepackage{amsmath}
\usepackage{graphicx, psfrag, epsf}
\usepackage{enumerate}
\usepackage{natbib}
\usepackage{url} 
\usepackage{xr}

\newcommand{\blind}{0}
 \newcommand{\noop}[1]{}

\makeatletter

\newcommand*{\addFileDependency}[1]{
\typeout{(#1)}
\@addtofilelist{#1}
\IfFileExists{#1}{}{\typeout{No file #1.}}
}\makeatother

\newcommand*{\myexternaldocument}[1]{%
\externaldocument{#1}%
\addFileDependency{#1.tex}%
\addFileDependency{#1.aux}%
}
\myexternaldocument{supplement}

\usepackage[margin=1in]{geometry}

\usepackage{setspace}
\usepackage{amssymb, amsfonts, amsthm, bm}
\usepackage{mathtools, mathrsfs}
\usepackage{relsize} 
\usepackage{hyperref}
\usepackage{comment}

\newtheorem{lemma}{Lemma}

\newtheorem{example}{Example}

\DeclareMathOperator{\N}{\mathrm{N}}
\DeclareMathOperator{\MN}{\mathrm{MN}}

\usepackage{color}

\begin{document}

	\if1\blind
	{
	\title{Calibration of imperfect geophysical models by multiple satellite interferograms with measurement bias}
 \vspace{-.15in}
			\date{}
		\maketitle
	}\fi
	
	\if0\blind
	{
\title{Calibration of imperfect geophysical models by multiple satellite interferograms with measurement bias}
 \vspace{-.2in}
		\author{  Mengyang Gu$^*$,  Kyle Anderson$^{**}$, and Erika McPhillips$^{*}$ \vspace{-.05in}\\
			$^{*}$ Department of Statistics and Applied Probability, UC Santa Barbara  \vspace{-.1in} \\
			$^{**}$ U. S. Geological Survey, Volcano Science Center
			}
			\date{}
		\maketitle
	} \fi
	\begin{abstract}
		\noindent 
		
		\thispagestyle{empty} 

\vspace{-.2in}

	Model calibration consists of using experimental or field data to estimate the unknown parameters of a mathematical model. The presence of model discrepancy and measurement bias in the data complicates this task. Satellite interferograms, for instance, are widely used for calibrating geophysical models in geological hazard quantification. In this work, we used satellite interferograms to relate ground deformation observations to the properties of the magma chamber at K\={\i}lauea Volcano in Hawai`i. We derived closed-form marginal likelihoods and implemented posterior sampling procedures  that simultaneously estimate the model discrepancy of physical models, and the measurement bias from the atmospheric error in satellite interferograms. 
	We found that model calibration by aggregating multiple interferograms and downsampling the pixels in the interferograms can reduce the computation complexity compared to calibration approaches based on multiple data sets. The conditions that lead to no loss of information from data aggregation and downsampling are studied. 
	Simulation  illustrates that both discrepancy and measurement bias can be estimated, and real applications demonstrate that modeling both effects helps obtain a  reliable estimation of a physical model's unobserved parameters and enhance its predictive accuracy.
   We implement the computational tools in the {\tt RobustCalibration} package available on CRAN. 

		
	\end{abstract}

	\newpage
\setcounter{page}{1} 


\section{Introduction}
\label{sec:intro}

Mathematical models are often used to describe various phenomena in science and engineering. To predict complex processes, one often first needs to estimate the unobserved parameters in the model using experimental observations or field data -- a process generally known as model calibration. Denote the mathematical model by $f^M(\mathbf x, \bm \theta)$, where $\mathbf x$ is a $p_x$-vector of observed input and $ \bm \theta$ is a  $p_{\theta}$-vector of calibration parameters, both assumed to be real-valued. The superscript `M' denotes the model. 
As the mathematical model may not represent reality perfectly, accurately estimating the difference between the mathematical model and reality can improve the predictive accuracy. 
In \cite{kennedy2001bayesian}, a Gaussian stochastic process (GaSP) defined on the observed input space, $\delta(\mathbf x)$, was proposed to model the discrepancy between the mathematical model and reality. Modeling the discrepancy using a GaSP was subsequently examined in a number of other applications \citep{bayarri2007framework,higdon2008computer,arendt2012quantification}.

In this work, we focus on spatially correlated patterns of measurement error from the data acquisition process, which we term \textit{measurement bias}. We study interferometric synthetic aperture radar (InSAR)  interferograms which, over the last 25 years, have made it possible to map deformation over broad swathes of the Earth's surface to sub-centimeter accuracy from space, revolutionizing scientists' understanding of Earth processes \citep{Massonnet1995,Burgmann2000,Pinel2014}. InSAR interferograms are most often obtained using data from orbiting microwave-band radar satellites. By interfering two radar images of the surface taken from a satellite at different times, changes in the radar phase track temporal changes in the position of the Earth's surface along the oblique line-of-sight (LoS) vector between the satellite and ground.  Because only fractional phase change can be measured directly, while the number of complete phase cycles between the satellite and ground is unknown, these images are wrapped by the radar's wavelength. Unwrapping an image by spatial integration of the phase gradient -- relative to a point believed to be non-deforming or having zero ground displacement -- yields relative LoS deformation in units of distance change \citep{Chen2001}.

Despite these advances, the interpretation of InSAR data is often greatly complicated by noise and bias. After removing the phase due to elevation of Earth's surface, the processed (or wrapped) phase for each pixel on the ground contains ground displacement, measurement bias, satellite orbital error, and look angle error.  Measurement biases, in particular, are known to strongly affect many interferograms,  caused most importantly by propagation delays due to atmospheric conditions, which yield spatially correlated noise which varies in time and space \citep[e.g.,][]{Zebker1997,Hooper2007,Gong2016}. Much work remains  for mitigating and characterizing these uncertainties. 

These observations motivate modeling the random measurement bias that can vary across different data sources due to  environmental conditions or  properties of devices, as well as a discrepancy function invariant across other data sources, to explain the difference between reality and model outcomes. The framework can be extended to  integrate different types of observations, such as satellite radar interferograms, GPS, and tiltmeter observations  \citep{anderson2019magma}.

	\begin{figure}[t!]
		\centering
			\includegraphics[width=.95\textwidth]{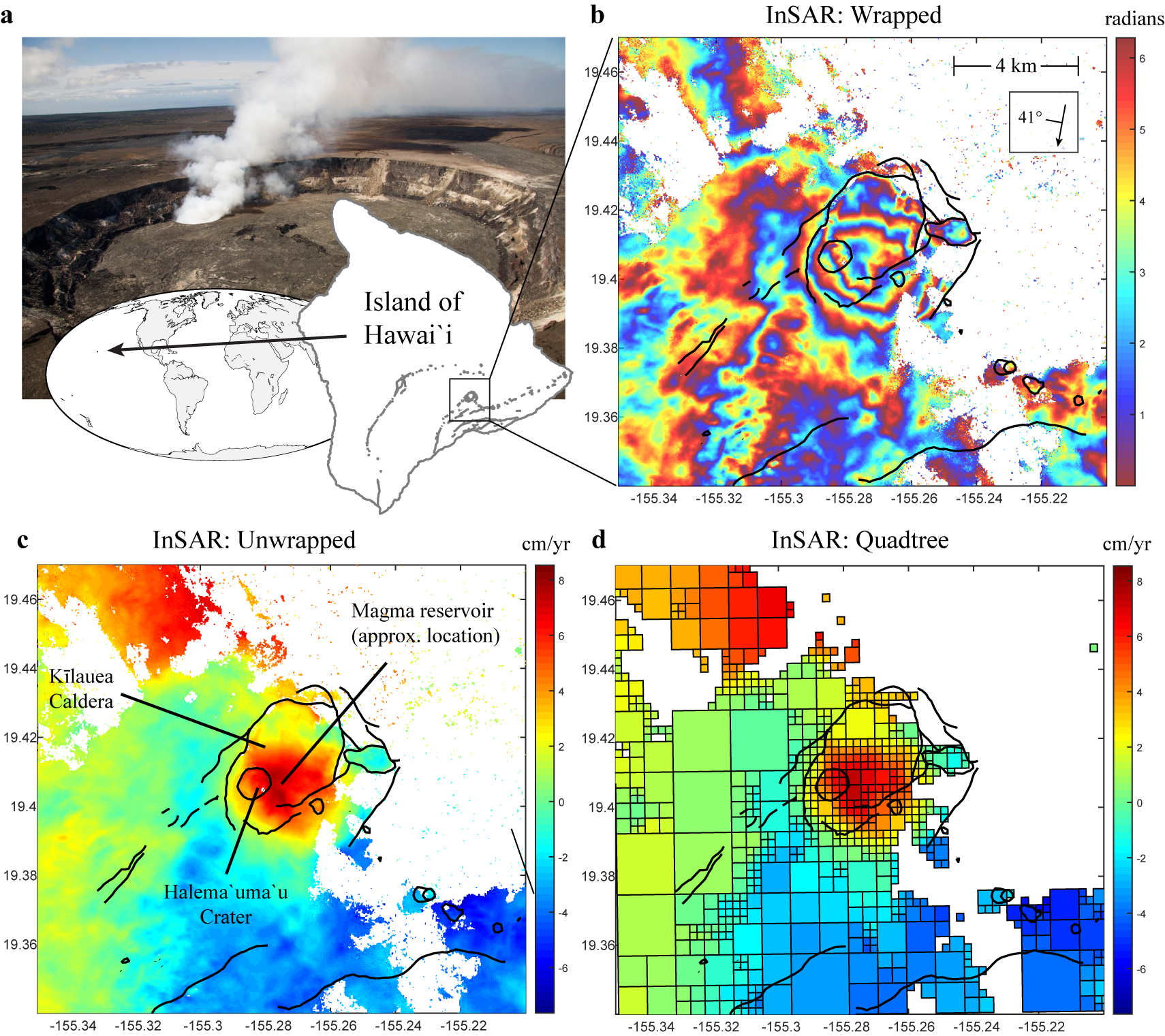}
		\caption{a) Overview maps showing the location of K\={\i}lauea Volcano on the Island of Hawai`i, with background photo of Halema`uma`u Crater at the volcano's summit roughly as it appeared during the time of this study 
		(USGS photo). (b) Wrapped InSAR interferogram from the COSMO-SkyMed satellite, spanning 20 Oct 2011 to 15 May 2012. The inset box shows the flight path of the satellite (arrow) and the downward look direction of the satellite at 41$^{\circ}$. White areas have no data due to radar decorrelation. Number of data points is around $1.5\times10^5$.
		(c) Same data as in (b), but unwrapped. (d) Quadtree-processed interferogram.
		Thick black lines in panels b-d show cliffs and other important topographic features at the volcano; the large elliptical feature is K\={\i}lauea Caldera.
		}
		\label{fig:Kyle_Kilauea}
	\end{figure}

We utilize data from K\={\i}lauea Volcano, one of the world's most active volcanoes. K\={\i}lauea is located on the Island of Hawai`i and erupted semi-continuously from 1983-2018. In 2018, a historically unprecedented rift eruption destroyed more than 700 homes and displaced thousands of residents \citep{neal20192018}. Figure~\ref{fig:Kyle_Kilauea} shows ground displacements at K\={\i}lauea from October 2011 to May 2012, as the volcano's summit inflated due to magma storage \citep{anderson2016bayesian}. This data was recorded from a satellite orbiting roughly north to south, which recorded LoS deformation 
along a vector oblique to the Earth's surface -- roughly east to west and downward at an angle of 41 degrees. This image thus resolves a combination of predominantly 
vertical and east-west ground deformation. 
In Figure~\ref{fig:Kyle_Kilauea}d, the image was subsampled  by the quadtree algorithm for computational efficiency \citep{Jonsson2002}.

Of the sources of uncertainty in InSAR observations, the spatially correlated atmospheric term is usually the most important; spatial and temporal changes of just 20\% in relative humidity can lead to errors of 10 cm in estimated ground deformation in some scenarios \citep{Zebker1997}. 
Figure \ref{fig:Kyle_Kilauea}b shows the wrapped InSAR phase at K\={\i}lauea Volcano. A ``bullseye'' pattern near the center 
is due to real ground deformation, while most of the remaining fringes are due to atmospheric conditions.

We highlight a few contributions of this study to address the main challenges involved in  calibrating models using data from multiple sources. First, although discrepancy functions have been studied extensively for model calibration, discrepancy and measurement bias are rarely studied together. In this work, we simultaneously model both discrepancy and measurement bias functions and estimate their effects using multiple InSAR interferograms. This approach allows one to estimate geophysical model discrepancy and measurement bias due to distinct atmospheric conditions recorded in each InSAR interferogram.  Furthermore, we derive the marginal likelihood and posterior distributions of the model parameters, which are useful for efficient posterior sampling. 
Third, InSAR interferograms can contain millions of pixels, so downsampling schemes such as the quadtree algorithm \citep{Simons2002b} are often applied, or multiple interferograms are averaged prior to modeling in order to reduce computational cost. However, the implications of modeling a single averaged interferogram rather than jointly modeling multiple individual interferograms has not been well-studied. Here we discuss the conditions under which these two approaches are equivalent and scenarios in which modeling individual data sets leads to more precise estimation.  Simulated studies demonstrate these findings and confirm that multiple data sources can estimate the shared discrepancy function and source-dependent measurement bias. 
Finally, the new method has been implemented in the {\tt RobustCalibration} package available on CRAN \citep{gu2022robustcalibration}. 





The rest of the paper is organized as follows. Section \ref{sec:multisources} introduces our approach that includes models of both the discrepancy function and measurement bias, as well as posterior sampling for Bayesian inference. Connections and differences between jointly modeling individual data sets and downsampled data, as well as different models of discrepancy functions, are also discussed.  
 Simulated and real examples comparing several models are given in Section \ref{sec:simulation} and Section \ref{sec:real}, respectively.
 We provide a short conclusion  in Section  \ref{sec:conclusion}. Lastly, supplementary materials contain extensive derivations of marginal likelihood functions, posterior distributions, theoretical results regarding the consistency of model calibration, and numerical comparisons of different  models of discrepancy functions and downsampling approaches. The code and data used in this article are publicly available: \url{https://github.com/UncertaintyQuantification/MultiCalibration}. 
\vspace{-.1in}
\section{Model calibration by multiple sources of data}
\label{sec:multisources}





 
Let us consider the model of the $l$th source of real-valued field  measurement, $y^F_l(\mathbf x)$, with superscript `F' meaning `field,' to calibrate an imperfect model with the observable input $\mathbf x \in \mathbb R^{p_x}$ and  unobservable calibration parameters $\bm \theta \in \mathbb R^{p_{\theta}}$, 
\begin{equation}
 y^F_l(\mathbf x)= f^M(\mathbf x, \bm \theta)+ \delta(\mathbf x)+ \delta_l(\mathbf x)+\mu_l  +\epsilon_{l}(\mathbf x).
\label{equ:main_model}
\end{equation}
Here, $\mu_l$, $\delta_l(\mathbf x)$, and $\epsilon_{l}(\mathbf x)\sim \N(0,\sigma^2_{0l})$ are the source-specific mean parameter, random measurement bias, and noise, respectively, for source $l=1,...,k$. $\delta (\mathbf x)$ is a discrepancy term between reality and the computer model shared across data sources, which is independent of computer models and thus only depends on the observed input, as advocated in \cite{kennedy2001bayesian}.  
In our application, $\mathbf x$ is the spatial coordinates of an InSAR interferogram, $\bm \theta$ are physical model parameters listed in Table \ref{tab:description}, and $k$ is the number of InSAR interferograms.  The satellite interferograms are taken at slightly different start and end dates to measure the ground deformation, and the true ground deformation is approximately the same in different interferograms. This motivates the inclusion of a discrepancy function $\delta (\mathbf x)$ shared across all sources.  On the other hand, the measurement bias $\delta_l(\mathbf x)$ is distinct in each interferogram, as the atmospheric conditions were different when each of the interferograms were taken. 

For a set of observable inputs  $\{\mathbf x_1,...,\mathbf x_n\}$, let us first assume that the marginal distributions of model discrepancy $\bm \delta=(\delta(\mathbf x_1),...,\delta(\mathbf x_n))^T$ and measurement bias $\bm \delta_l=(\delta_l(\mathbf x_1),...,\delta_l(\mathbf x_n))^T$ follow  multivariate normal distributions: $ \bm \delta \sim \MN(\mathbf 0, \tau^2 \mathbf R)$ and $\bm \delta_l \sim \MN(\mathbf 0, \sigma^2_l\mathbf R_l)$, respectively. The covariance matrix of the discrepancy is denoted as $\tau^2 \mathbf R$, which contains the variance parameter $\tau^2$; the $(i,j)$th entry of the correlation matrix $\mathbf R$ is parameterized by a kernel function $K(\mathbf x_i,\mathbf x_j)$, while the $(i,j)$th entry of covariance matrix of the $l$th measurement bias is $\sigma^2_lK_l(\mathbf x_i,\mathbf x_j)$. Here, $\sigma^2_l$ is the variance parameter for  $l=1,..,k$.
We postpone the discussion of kernel functions and other models of discrepancy functions to Section \ref{subsec:model_discrepancy_bias}. 

In model (\ref{equ:main_model}), the physical reality, denoted as $y^R(\mathbf x)$ at any coordinate $\mathbf x$, can be expressed as a summation of the mathematical model and discrepancy function, i.e.,
$y^R(\mathbf x)=f^M(\mathbf x, \bm \theta)+ \delta(\mathbf x)$, which follows the framework in \cite{kennedy2001bayesian}. The innovation in (\ref{equ:main_model}) is to explicitly model the measurement bias (spatially correlated pattern) contained in different sources of observations. Here we have two goals. The first goal is to estimate the calibration parameters, discrepancy function, and measurement bias. The second goal is to predict physical reality by combining the calibrated physical model with the discrepancy function. 


InSAR measures ground displacements relative to a point assumed to be non-deforming, i.e., a spatial location assumed to have zero ground deformation, which introduces uncertainty. We therefore include an unknown mean parameter $\mu_l$ for each interferogram $l$ in model (\ref{equ:main_model}) and estimate it using data. InSAR images may also contain long-wavelength ``ramp'' artifacts due to errors in satellite orbits, which may be corrected independently (for instance, using data from GPS sensors) \citep{Simons2007} or by estimation of linear or quadratic ramp parameters together with geophysical model parameters. For our case study, however, the geographic area of interest is relatively small, so we neglect these errors.

 The closed-form marginal distributions and predictive distributions of the discrepancy, measurement bias, and the reality of model (\ref{equ:main_model}) are given in Section 
 S2 
 of the supplementary materials. Since the computational complexity of each evaluation of the likelihood increases linearly to $k$, images are often averaged before modeling in geoscience studies to reduce computational cost. We first study the difference between these two ways of inference. 

 



\subsection{Model equivalence based on aggregated data and full data}
\label{subsec:stack_image}



First, let us consider the data with no correlated measurement bias $\delta_l(.)$, that is  $y^F_l(\mathbf x)=y^R(\mathbf x)+\epsilon_l(\mathbf x)$. Here, $y^R(\mathbf x)$ is the  unknown reality.   The  data can be modeled below 
\begin{equation}
y^F_{l}(\mathbf x)=f^{M}(\mathbf x, \bm \theta)+\delta(\mathbf x)+\mu+\epsilon_l(\mathbf x), 
\label{equ:ind}
\end{equation}
where the independent noise follows $\epsilon_l(\mathbf x) {\ \sim \ } N(0,\sigma^2_0)$ for each $\mathbf x$, and  $l=1,...,k$. 
We include the mean parameter $\mu$ here for  scenarios where the mean of the reality is not directly modeled in the physical model $f^{M}$. 
Model (\ref{equ:ind}) has been widely used in calibration of repeated experimental responses \citep{bayarri2007framework,arendt2012improving}.


Denote $\bar y^F(\mathbf x)=\sum^k_{l=1}y^F_{l}(\mathbf x)/k$ as the average value of the field data at the input $\mathbf x$.  When (\ref{equ:ind}) is assumed,  the model of the aggregated data follows
\begin{equation}
\bar y^F(\mathbf x)=f^{M}(\mathbf x, \bm \theta)+  \delta(\mathbf x)+ \mu+\bar \epsilon(\mathbf x), 
\label{equ:stack}
\end{equation}
where the noise independently follows $\bar \epsilon(\mathbf x) \sim  N(0,  \sigma^2_0/k)$ for each $\mathbf x$. In our real application, the InSAR interferograms are aligned on the same spatial coordinates $\{\mathbf x_1,...,\mathbf x_n\}$. Since the number of spatial coordinates is large, the uncertainty in aligning the spatial coordinates of InSAR interferograms is approximately negligible.  
Also, denote $\mathbf y^F_l=(y_l^F(\mathbf x_1),...,y^F_l(\mathbf x_n))^T$ as the observations in  source $l$ and  $\bar{\mathbf y}^F=(\sum^k_{l=1}y^F_{l}(\mathbf x_1)/k,...,\sum^k_{l=1}y^F_{l}(\mathbf x_n)/k)^T$ as the aggregated data. In Lemma \ref{lemma:lik_aggregate_connection} below, we show the logarithm of the likelihood of the full data and the reduced data only differs by a constant relevant to the variance of the noise. The proof is given in Section S1 in the supplementary materials.



 




 
 
 \begin{lemma}
 Integrating out $\bm \delta \sim \mbox{MN}(\mathbf 0, \tau^2 \mathbf R)$, the natural logarithm of the marginal likelihood in  model (\ref{equ:ind}) follows 
   \begin{equation}
 \ell(\bm \theta, \mu, \sigma^2_0, \tau, \mathbf R )=c_{\sigma^2_0}+ \bar \ell(\bm \theta, \mu, \sigma^2_0, \tau, \mathbf R),
 \label{equ:likelihood_ratio}
 \end{equation}
 where $c_{\sigma^2_0}=-\frac{n(k-1)}{2}\log(2\pi\sigma^2_0)-\frac{n}{2}\log(k)-\frac{\sum^k_{l=1}\sum^n_{i=1}( y^F_l(\mathbf x_i)-\bar{ y}^F(\mathbf x_i))^2 }{2\sigma^2_0}$ and $\bar \ell(\bm \theta, \mu, \sigma^2_0, \tau, \mathbf R)$ is the natural logarithm of the marginal likelihood model from (\ref{equ:stack}): 
  \[\bar \ell(\bm \theta, \mu, \sigma^2_0, \tau, \mathbf R )=-\frac{n}{2}\log(2\pi)
  -\frac{1}{2}\log\left|\tilde {\bm  \Sigma}\right|- \frac{(\bar{\mathbf y}^F-\mu \mathbf 1 -\mathbf f^M_{\bm \theta})^T \tilde {\bm  \Sigma}^{-1} (\bar{\mathbf y}^F-\mu \mathbf 1 -\mathbf f^M_{\bm \theta}) }{2},\]
  with $\tilde {\bm  \Sigma}= \tau^2 \mathbf R+\frac{\sigma^2_0}{k}\mathbf I_n$ and  $\mathbf f^M_{\bm \theta}=(f^M_{\bm \theta}(\mathbf x_1, \bm \theta),...,f^M_{\bm \theta}(\mathbf x_n, \bm \theta))^T$.
\label{lemma:lik_aggregate_connection}
 \end{lemma}
\vspace{-.2in}
When the variance of the noise $\sigma^2_0$ is known, 
equation (\ref{equ:likelihood_ratio}) implies that the estimation of calibration parameters, mean parameters, and discrepancy function based on the aggregated data and full data  is  the same, 
as the aggregated data $\bar{\mathbf y}^F$ is the sufficient statistics  of these parameters \citep{casella2021statistical}. 
When  $\sigma^2_0$ is unknown, the sufficient statistics are $\bar{\mathbf y}^F$ and $ s^2$, where $s^2=\sum^k_{l=1}\sum^n_{i=1}( y^F_l(\mathbf x_i)-\bar{ y}^F(\mathbf x_i))^2$. This result was previously discussed in \cite{bayarri2007framework}. However,  the  efficiency of the estimators based on full and aggregated data was  not compared. For instance, if the reality $y^R(\cdot)$ is a deterministic function, the usual unbiased estimator of $\sigma^2_0$ based on the full data is the sample variance $s^2/(n(k-1))$. Also, the variance when estimating $\sigma^2_0$ would be $2\sigma^4_0/(n(k-1))$. For aggregated data modeled in (\ref{equ:stack}), even if the reality $y^R(\cdot)$ is known, the estimator of the variance $\sigma^2_0$ based on the sample variance is $k \sum^n_{i=1}(\bar{y}^F(\mathbf x_i)-  y^R(\mathbf x_i))^2/(n-1)$, which has a variance of $2n\sigma^4_0/(n-1)^2$. This is larger than the sample variance  $2\sigma^4_0/(n(k-1))$  based on full data. Since reality $y^R(\cdot)$ is unknown, the variance of the estimator of  $\sigma^2_0$ based on the aggregated data typically becomes even larger. Thus, modeling aggregated data in (\ref{equ:stack}) is less efficient in estimating  $\sigma^2_0$, compared to modeling the full data in (\ref{equ:ind}). 

Second, for model (\ref{equ:main_model}) that contains the measurement bias, we have a similar result as Lemma \ref{lemma:lik_aggregate_connection}. The logarithm of the likelihood of  model (\ref{equ:main_model})  can be decomposed into two parts as in equation (\ref{equ:likelihood_ratio}), where the first part contains individual data vectors $\mathbf y_1,...,\mathbf y_k$, a weighted mean data vector $\mathbf {\bar y}_{w}=\sum^k_{l=1}(\mathbf y_l-\bm \delta_l-\mu_l \mathbf 1_n)/(k\sqrt{\sigma_{0l}})$, and variances of the noises $\sigma^2_{01},...\sigma^2_{0k}$. The second part contains aggregated data $\mathbf {\bar y}_{w}$, calibration parameters $\bm \theta$, and covariance parameters. Thus, when the variances of the noises are known,  parameter estimation based on the aggregated data $\mathbf {\bar y}_{w}$ and individual data sets is the same when the measurement bias is included. However, note that $\mathbf {\bar y}_{w}$  is different from the average $\mathbf {\bar y}$ and is generally not observable.  In our real application, this result indicates that using the averaged interferogram can lead to the loss of information if each interferogram contains distinct atmospheric errors, mean parameters, or unknown noise variance. Simulated studies in Section \ref{sec:simulation} further confirm this result. 

\subsection{Downsampling satellite interferograms}
\label{subsec:downsample}
A single InSAR image is often composed of hundreds of thousands of pixels. Even for very simple geophysical models, the expense of computing deformation at all these points can  be prohibitive, and subsampling techniques are typically employed. One approach is to uniformly sample a subset of pixels  
for  calibration and prediction \citep{Pritchard2002}.    As a result, posterior distributions of  calibration parameters are often stable, with responses at only a few hundred pixels \citep{gu2018sgasp}.


Another method of downsampling is the quadtree algorithm, in which one computes the average (or median) of groups of pixels (``boxes''), whose sizes are based on gradients in the image 
\citep{Simons2002b}, the resolution of the forward model \citep{Lohman2005}, or 
both \citep{Wang2014}. The quadtree algorithm clusters the pixels in smaller boxes for regions with rapid changes in pixel values, while areas with less change are clustered in larger boxes. These algorithms have become widely used for modeling InSAR data   \citep[e.g.,][]{Montgomery-Brown2015a,anderson2019magma}. The quadtree algorithm reduces the data from around a million pixels to a few hundred boxes, effectively reducing the number of observations by more than a thousand times. The quadtree algorithm 
may be considered a type of supervised data reduction method where the output values on the pixels are used for clustering. Other supervised algorithms for data reduction,  such as those discussed in \cite{joseph2021supervised}, could also be useful.  


Because the boxes in a quadtree-processed image, such as in Figure \ref{fig:Kyle_Kilauea}d, are computed from different numbers of pixels, in a calibration problem it is therefore important to consider the boxes' size 
\citep{Simons2002b,Lohman2005}. However, this seems to have been overlooked in many previous studies using quadtree-processed InSAR data for model calibration and prediction. 



Suppose the $l$th quadtree-processed image is composed of $J_l$ boxes, each box computed from $n_{j,l}$ pixels, for $j=1,...,J_l$ and $l=1,...,k$. Denote the $l$th quadtree-processed image by  $\mathbf y^{F,Q}_l:=\{y^{F,Q}_{1,l},...,y^{F,Q}_{J_l,l}\}$, where $y^{F,Q}_{j,l}$ is the average of the pixels of the $j$th quartree box for $j=1,...,J_l$. 
 Let $\mu^Q_l$, $f^{M,Q}_{j,l}( \bm \theta)$,  $\delta^Q_{j,l}$, and $\delta^Q_{j,l}$ denote the corresponding mean parameter, outputs of the mathematical model, discrepancy function, and measurement bias function evaluated at the centroid of the $j$th quadtree box at the $l$th  quadtree-processed image. Consider the model of the quadtree-processed image $l$:
\begin{equation}
y^{F,Q}_{j,l}=f^{M,Q}_{j,l}( \bm \theta)+\mu^Q_l+\delta^Q_{j,l}+\delta_{j,l}+ \epsilon^Q_{j,l},
\label{equ:y_F_quadtree_normal}
\end{equation}
where  $\epsilon^Q_{j,l} $ is a zero-mean Gaussian noise with variance  $\frac{\sigma^2_0}{\omega_{j,l}}$ and  $\omega_{j,l}$ is the weight for the $j$th image and $l$th source of data. Here we suppress the notation of spatial inputs as the point process data are compressed to areal data, and  the correlation structure is defined between a finite set of areal units.   
Denote $S_{j,l}$ as the index set of pixels, where the pixels in this set belong to the $j$th quadtree box in the $l$th data source. 
The likelihood of model (\ref{equ:y_F_quadtree_normal}) of the quadtree-processed image is the same as that for model  (\ref{equ:main_model}) of the original image if
\begin{align} 
\begin{split}
&\sum_{i\in {S_{j,l}} } (y^F_l(\mathbf x_i)-f^{M}(\mathbf x_i, \bm \theta)-\delta(\mathbf x_i)-  \delta_l(\mathbf x_i)-\mu_l )^2  \\
&\quad \quad \quad =  \omega_{j,l} (y^{F,Q}_{j,l}-f^{M,Q}_{j,l}(\bm \theta)-\delta^Q_{j}-\delta^Q_{j,l}-\mu^Q_l )^2.
\end{split}
\label{equ:cond_quad}
\end{align}
Other physical measurements may be useful for estimating discrepancy  and measurement bias; however, in other cases, one may not know the discrepancy  and measurement bias functions apriori. Some weighting schemes utilize the estimated covariance structure of the InSAR data \citep{Lohman2005}.  
 In this work, we follow  \cite{Simons2002b} and \cite{anderson2016bayesian} by letting the weight of each quadtree box be proportional to the number of pixels in the box, i.e., $\omega_{j,l} \propto n_{j,l}$ for $j=1,...,J_l$. A quadtree box with a larger size has a larger weight because it is averaged with more pixels. 

\subsection{Statistical models of discrepancy and measurement bias}
\label{subsec:model_discrepancy_bias}
We discuss specific discrepancy and measurement bias functions in this subsection. The discrepancy function is often modeled as a GaSP \citep{kennedy2001bayesian}:
 \begin{equation}
 \delta(\cdot)\sim \mbox{GaSP}(0, \tau^2 K(\cdot,\cdot)),
 \label{equ:delta_GaSP}
 \end{equation}
 where $\tau^2 K(\cdot,\cdot)$ is a covariance function with variance $\tau^2$.
 The identifiability issue, however, has been widely observed in modeling spatially correlated data, where the spatial random effect was confounded with a  linear fixed effect, i.e., $f^M(\mathbf x, \bm\theta)$ being a linear model of $\bm \theta$  \citep{reich2006effects,hodges2010adding,zhang2004inconsistent}. \cite{wang2020prediction}, for example, show that the variance of generalized least squares estimator of the linear coefficients is bounded above zero under infill asymptotics.    The non-identifiability of the calibration parameters was also recently observed when the discrepancy function was modeled by a GaSP \citep{arendt2012quantification,tuo2015efficient,tuo2016theoretical,plumlee2016bayesian,wong2017frequentist}, where the calibrated physical models can be far away from  reality in terms of $L_2$ distance.

 
 Previous studies \citep{bayarri2007framework,arendt2012improving} suggest that repeated observations help identifiability as  estimation accuracy of  the variance of the data improves, which coincides with our discussion in Section \ref{subsec:stack_image}. 
 However, when the discrepancy is sampled from the true model, the MLE of the calibration parameter may not be consistent even if we have infinite repeated measurements, i.e., $k\to \infty$.  An example is provided in Example S1
 in the supplementary materials to illustrate this finding. 
 This means that the repeated measurements are helpful in estimating the variance of the noise in the data, yet it cannot solve the identifiability issue. 


When we model the discrepancy function by a GaSP with commonly used covariance functions, such as power exponential covariance or Mat{\'e}rn covariance \citep{rasmussen2006gaussian},  the $L_2$ loss between the mathematical model and reality is $L_2(\bm \theta)= \int_{\mathbf x\in \mathcal X} (y^R(\mathbf x)-f^M(\mathbf x, \bm \theta) )^2 d \mathbf x=  \int_{\mathbf x\in \mathcal X} \delta^2(\mathbf x)d \mathbf x$. $L_2(\bm \theta)$ is a random variable whose measure is induced by the covariance  function of the GaSP. The distribution of $L_2(\bm \theta)$  can have a substantial probability mass at a large $L_2$ loss when the correlation in the data is large. In \cite{gu2018sgasp}, this random $L_2$ loss is scaled to have more probability mass near zero. The construction of the discretized S-GaSP is summarized in S4
 in the supplementary materials.

Starting from a GaSP model with any reasonable covariance function $\tau^2 K(\cdot, \cdot)$, and integrating out $Z$, the marginal distribution of the discretized GaSP $\bm \delta_z:=(\delta_z(\mathbf x_1),...,\delta_z(\mathbf x_n))^T$ follows a multivariate normal distribution with the following transformed covariance matrix:
	\begin{equation} 
		\bm \delta_z \mid \tau, \mathbf R_z   \sim \text{MN}( \bm 0,   \tau^2 \mathbf R_z ), 
		\label{equ:delta_z_marginal_approx}
	\end{equation}
	where $\mathbf R_z=\left(\mathbf R^{-1} +\frac{\lambda_z}{n}\mathbf I_n\right)^{-1}$ 	and the $(i,\, j)$th term of $\mathbf R$ is $K(\mathbf x_i, \mathbf x_j)$  with range parameters $\bm \gamma=(\gamma_1,...,\gamma_{p_x})^T$. Note that $\mathbf R_z$ is different from a covariance matrix with a nugget parameter. 




A larger $\lambda_z$ assigns more prior probability on the smaller sum of squares of the discrepancy function. Under some regularity conditions, a suitable choice of  $\lambda_z$  guarantees the predictive distribution in the S-GaSP model converges to the reality as fast as in the GaSP model. The estimation of the calibration parameters in the S-GaSP model minimizes the $L_2$ loss between the reality and mathematical model when the sample size goes to infinity. 
In numerical examples, we let $\lambda_z=C \sqrt{n}$, with $C=100$, which guarantees two  convergence  properties under common regularity conditions \citep{gu2018theoretical}.

Furthermore, the measurement biases are spatially correlated, which can be modeled as a spatial random effect 
 via a GaSP, 
 \begin{equation}
 \delta_l(\cdot)\sim \mbox{GaSP}(0, \sigma^2_l K_l(\cdot,\cdot)),
 \label{equ:delta_l_GaSP}
 \end{equation}
where $K_l(\cdot,\cdot)$ is the kernel function for $l=1,...,k$. For any $\{\mathbf x_1,...,\mathbf x_n \}$, the marginal distribution of $\bm \delta_l$ follows a multivariate normal distribution with covariance $\sigma^2_l \mathbf R_l$, for $l=1,...,k$, with the $(i,j)$th term of $\mathbf R_l$ being $K_l(\mathbf x_i, \mathbf x_j)$. 
For any inputs $\mathbf x_a:=(x_{a1},...,x_{ap})$ and $ \mathbf x_b:=(x_{b1},...,x_{bp})$, we assume a product covariance  \citep{bayarri2007framework} 
\begin{equation}
K_l(\mathbf x_a, \mathbf x_b)= \prod_{t=1}^{p_x }K_{l,t}( x_{at},  x_{bt}),
\label{equ:K_product}
\end{equation}
 where each $K_{l,t}(\cdot,\cdot)$ is a one-dimensional kernel function for the correlation between the $t^{th}$ coordinate of any two inputs for the source $l$, $l=1,...,k$.  Denote $d_t=|x_{at}- x_{bt}|$. 
 One popular choice is the Mat{\'e}rn correlation, which  has a closed form expression  with the roughness parameter $\alpha=(2k+1)/2$ for $k \in \mathbb N$. That is, the Mat{\'e}rn correlation function with  $\alpha = 5/2$  \citep{handcock1993bayesian} has the expression below, 
\begin{equation}
	K_{l,t}(d_t)=\left(1+\frac{\sqrt{5}d_t}{\gamma_{l,t}}+\frac{5d_t^2}{3\gamma_{l,t}^2}\right)\exp\left(-\frac{\sqrt{5}d_t}{\gamma_{l,t}}\right).
	\label{equ:Matern}
\end{equation}
A desirable feature of the Mat{\'e}rn correlation is that the sample path of the process is $\lfloor \alpha\rfloor$ differentiable. Concerning the present scientific goal, we also note that previous works have argued that Mat{\'e}rn correlation functions are suitable for modeling atmospheric noise in InSAR data \citep{Knospe2010}. However, we do not limit ourselves to any specific correlation function, and the methods discussed in this work apply to all such functions.

\subsection{Prior distributions and posterior sampling}
\label{sec:param_est}




We assume the calibration model follows (\ref{equ:main_model}) with data sets from multiple sources, where  users can choose either GaSP or S-GaSP to model the discrepancy function. The marginal likelihood  and predictive distributions are provided in Section S2
 in the supplementary materials.  
Here, the parameters contain the calibration parameters, mean parameters, range and variance parameters of the discrepancy and measurement bias. 
For  computational purposes, we define the nugget parameter $\eta_{l}:=\sigma^2_{0l}/\sigma^2_l$, 
  the inverse range parameters $\beta_t=1/\gamma_t$, and  $\beta_{l,t}=1/\gamma_{l,t}$, for $l=1,...,k$ and $t=1,...,p_x$.

We assume the prior of the parameters below:
\begin{equation}
\pi(\bm \theta,	\bm \mu, \bm \beta_{1:k}, \bm \eta, \bm \sigma^2, 	\bm \beta,  \tau^2)\propto\frac{\pi(\bm \theta)\pi(\bm \beta) }{\tau^2}\prod^k_{l=1}\left\{\frac{\pi(\bm \beta_l, \eta_l)}{\sigma^2_l}\right\},
\label{equ:prior} 
\end{equation}
where the prior of the calibration parameters $\bm \theta$ often depends on experts' knowledge, as the calibration parameters have scientific meanings. For the simulated examples and the real example of calibrating the geophysical model of K\={\i}lauea Volcano, we assume $\pi(\bm \theta)$ is a uniform distribution on the domain of the calibration parameters. The mean and scale parameters are assigned a usual location-scale prior in (\ref{equ:prior}), i.e., $\pi(\tau^2)\propto 1/\tau^2$ and $\pi(\mu_l, \sigma^2_l)\propto 1/\sigma^2_l$, for $l=1,...,k$. Furthermore, we assume a jointly robust  prior for the range and nugget parameters in the measurement bias functions and discrepancy function \citep{gu2018jointly}.

We have implemented the posterior sampling for Bayesian model calibration and prediction using single or multiple data sets in the {\tt RobustCalibration} R package, available on CRAN. Users can specify the model with or without the measurement bias. Both the GaSP and discretized S-GaSP discrepancy models are implemented for users to choose as well. Furthermore, the geophysical model used in this study is computationally inexpensive, but that is often not the case  \citep[e.g.,][]{Anderson2011}. In such cases, a statistical emulator can be used to approximate the expensive computer model. 
  The  GaSP emulator from the {\tt RobustGaSP} package is  implemented in the {\tt RobustCalibration} package for emulating costly computer  models with scalar or vectorized outputs. These tools can be used for different  applications of model calibration and prediction.
  

\section{Simulated examples}
\label{sec:simulation}

We study simulated examples of model calibration and prediction in this section. An example comparing the GaSP and S-GaSP models of the discrepancy function is provided in Section S5 
of the supplementary materials, where the reality is from a deterministic function.  
Here we discuss an example where the discrepancy function and measurement bias are sampled from GaSPs, and we compare model calibration based on full and aggregated data. 
 

 \begin{example}
\label{eg:simulation_stack_bias}
Assume data is sampled from model (\ref{equ:main_model}), where  $f^M(x,  \theta)=\sin(\theta x)$, with $\theta=\pi/2$. The model discrepancy and measurement bias are  sampled from $ \bm \delta \sim \MN(\mathbf 0, \tau^2 \mathbf R)$ and $\bm \delta_l \sim \MN(\mathbf 0, \sigma^2_l\mathbf R_l)$, where $\tau=0.2$ and $\sigma_l=0.4+0.4(l-1)/(k-1)$. $\mathbf R$ and $\mathbf R_{l}$ are both parameterized by the Mat{\'e}rn kernel in (\ref{equ:Matern}) with $\gamma=0.1$ and $\gamma_l=0.02$,  respectively. The standard deviation of the noise is $\sigma_{0l}=0.05$, and the observations are equally spaced at $x_i \in [0,1]$ for $i=1,...,n$ and $l=1,...,k$.  We let $n=100$ and implement $N=200$ experiments at three configurations with $k=5$, $k=10$, and $k=15$. 
\end{example}

We use Example \ref{eg:simulation_stack_bias}  to illustrate that modeling the individual data is more accurate than the aggregated data when measurement bias exists. 
We record the performance of three models. The first and second approaches are the GaSP calibration and S-GaSP calibration based on the full data, where the discrepancy function is modeled by  GaSP in (\ref{equ:delta_GaSP}) and the discretized S-GaSP model with the marginal distribution in  (\ref{equ:delta_z_marginal_approx}), respectively.
Here the GaSP calibration model is the true sampling model,  as the discrepancy is sampled from a GaSP. We include S-GaSP calibration model to illustrate that the S-GaSP model has comparable performance to the GaSP model even if the true discrepancy function is sampled from  the GaSP calibration model. 
Also included is  the GaSP calibration using aggregated data, i.e., the averages of all sources of data. We draw $20,000$ posterior samples of the parameters for each approach, with the first $4,000$ posterior samples  used as the burn-in samples. To reduce storage space, we thin posterior samples by ten times.



 	  \begin{figure}[t!]
		\begin{tabular}{cc}
		   \includegraphics[height=0.35\textwidth,width=.48\textwidth]{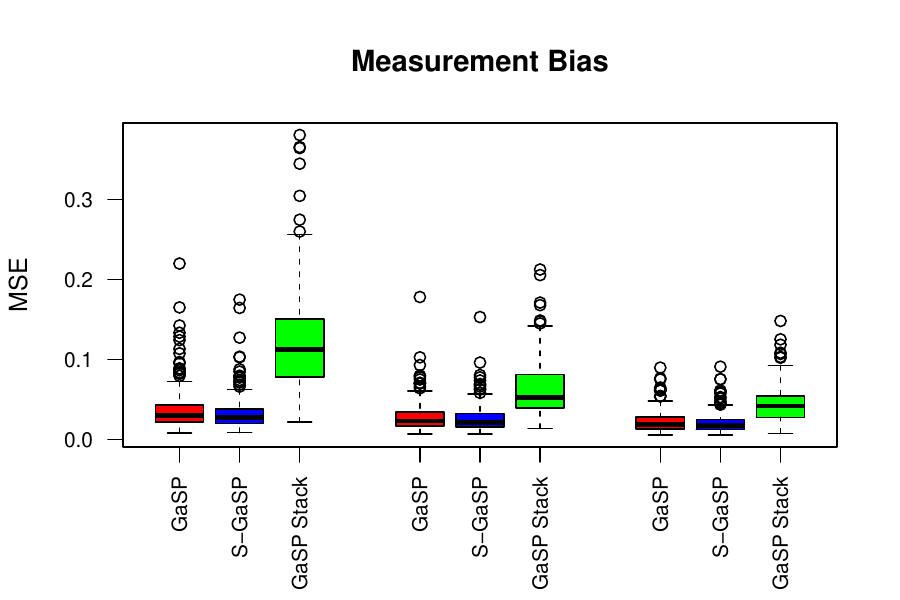}
		   \includegraphics[height=0.35\textwidth,width=.48\textwidth]{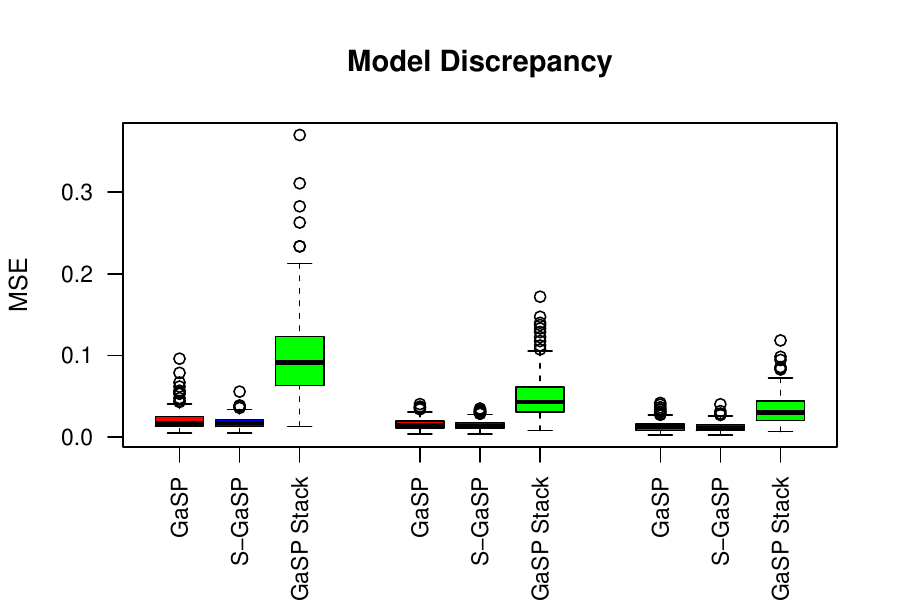}	\vspace{-.18in}\\ 
		   \includegraphics[height=0.35\textwidth,width=.48\textwidth]{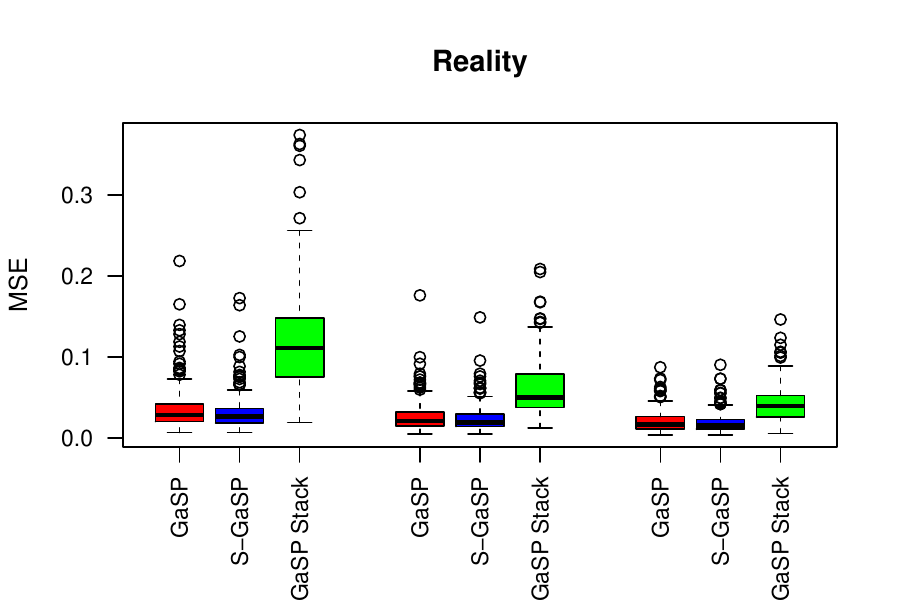}
		   \includegraphics[height=0.35\textwidth,width=.48\textwidth]{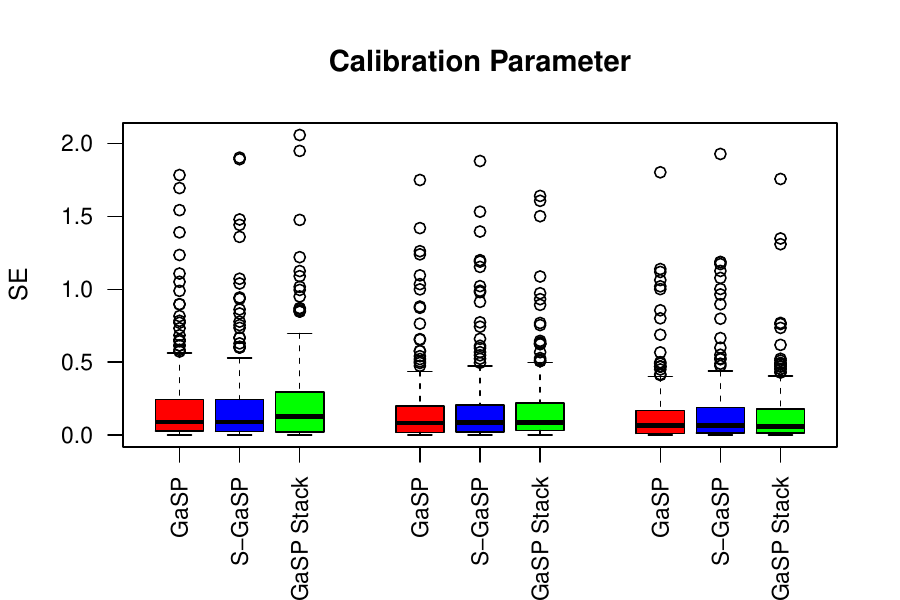}	\vspace{-.1in}
		\end{tabular}
		\caption{MSE of measurement bias  discrepancy function, reality and SE of the  calibration parameters for Example \ref{eg:simulation_stack_bias}.  
		In each panel, the first three boxes, the middle three boxes, and the right three boxes are the results when $k=5$, $k=10$, and $k=15$, respectively.  The MSE of GaSP calibration, S-GaSP calibration based on the full data, and the GaSP calibration based on the aggregated data (GaSP Stack) are colored as red, blue, and green. }
		\label{fig:MSE_simulation}
	\end{figure}


The mean square error (MSE) of measurement bias,  discrepancy functions,  reality, and the squared error (SE) of the calibration parameter of each experiment using different approaches for Example \ref{eg:simulation_stack_bias} are shown in Figure \ref{fig:MSE_simulation}.   First, even though the data is sampled from the GaSP calibration model, the MSEs of the S-GaSP calibration and GaSP calibration  are similar in estimation. We are not trying to show S-GaSP calibration can outperform GaSP calibration in this example, as the GaSP calibration model is the true sampling model. 
We include a simulated study in Example S2 
in the supplementary materials, to illustrate the identifiability problem of GaSP prior of the discrepancy function and the better performance of the S-GaSP calibration model.  Second, both methods based on the full data are better than the GaSP calibration based on the aggregated data, as averaging different sources of data causes loss of information due to the presence of the measurement bias and the unknown variance of the noise,  discussed  in Section \ref{subsec:stack_image}.  The estimation of the calibration parameter by the three methods is similar. When the number of sources of data increases, all methods become more accurate in estimation. The decrease of SEs of the calibration parameter is small when the number of sources of the observations increases because of the identifiability issue as the true discrepancy function is sampled from a GaSP. 

	  \begin{figure}[t!]
		\begin{tabular}{cc}
		   \includegraphics[height=0.37\textwidth,width=.48\textwidth]{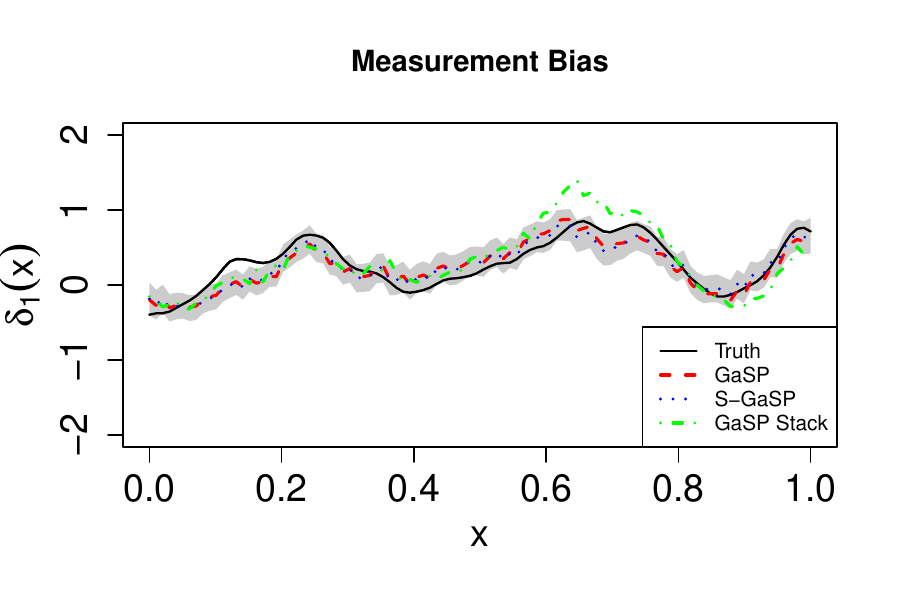}
		   \includegraphics[height=0.37\textwidth,width=.48\textwidth]{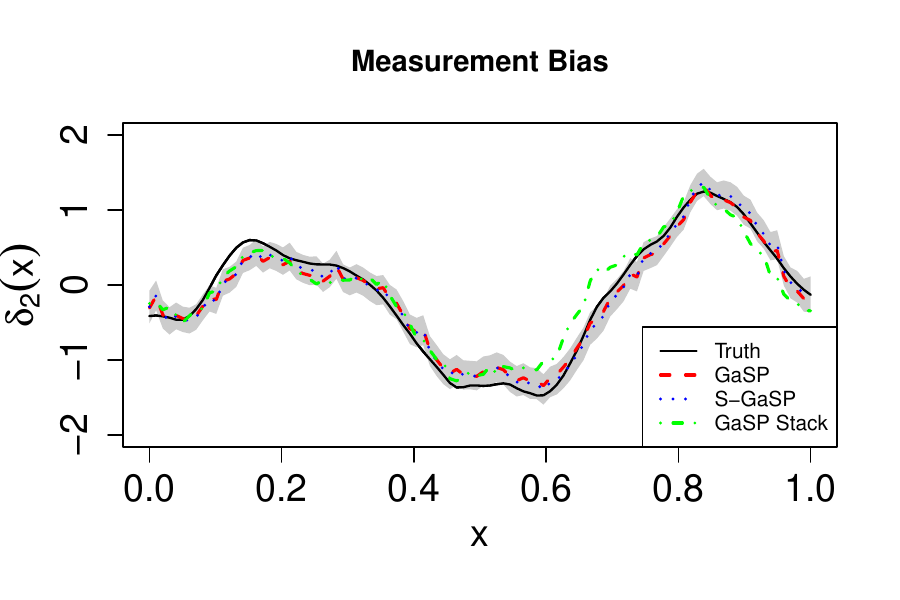}		\vspace{-.25in}\\ 
		   \includegraphics[height=0.37\textwidth,width=.48\textwidth]{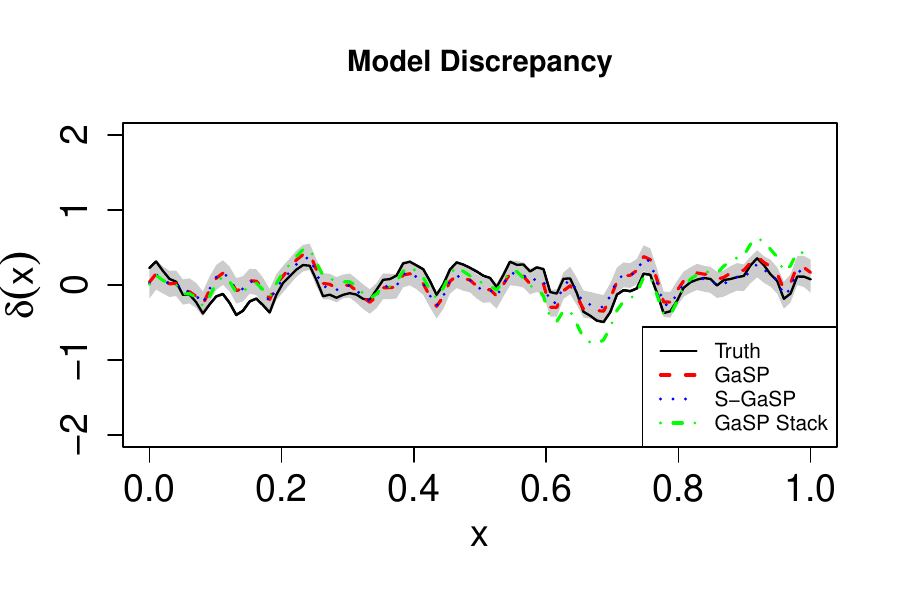}	
		   \includegraphics[height=0.37\textwidth,width=.48\textwidth]{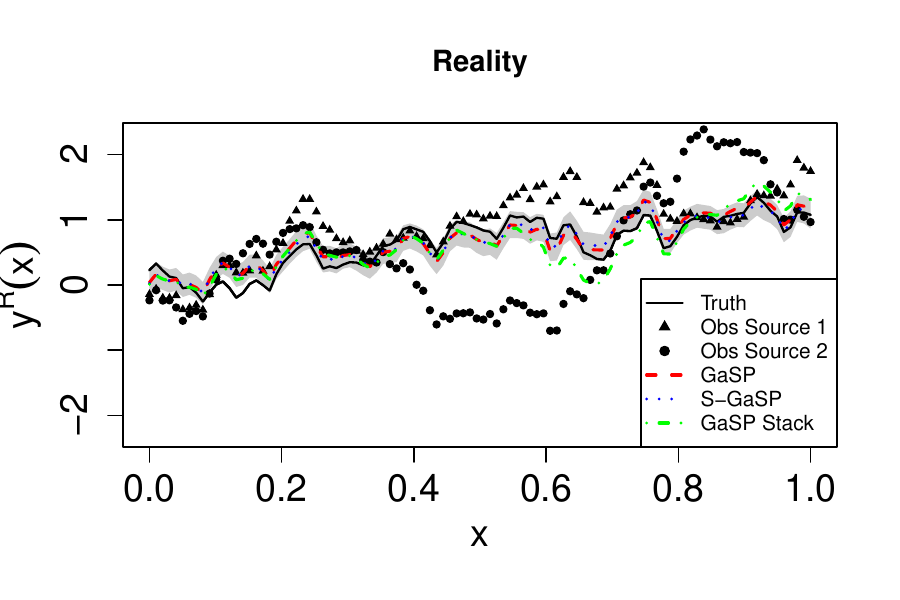}	\vspace{-.2in}
		\end{tabular}
		\caption{ The measurement bias in the first two sources, the model discrepancy function, and the reality in the first experiment of Example \ref{eg:simulation_stack_bias} when $k=10$ are graphed in the upper panels, lower left panel, and lower right panel, respectively. The truth and the estimation by the GaSP and S-GaSP calibrations based on the full data are graphed as the black solid lines, the red dashed lines, and blue dotted lines, respectively.  The estimation of the GaSP calibration based on aggregated data, denoted as GaSP Stack, is graphed as the green dotted and dashed lines. The 95\% posterior credible intervals from the S-GaSP calibration is graphed as the shaded area. The observations from the first two sources are graphed as black   triangles and dots, respectively, in the lower right panel. The black, red,  and blue lines almost overlap in all panels, indicating higher accuracy  based on individual data than the aggregated data, when the observations contain measurement bias.   }
		\label{fig:plot_simulation}
	\end{figure}
	
	 We graphed the measurement bias, the model discrepancy, reality, and their estimations in the first simulated experiment of Example \ref{eg:simulation_stack_bias} with  $k=10$ in Figure \ref{fig:plot_simulation}.  All methods seem to capture the patterns of the measurement bias, model discrepancy, and reality. The estimation conducted by the GaSP and S-GaSP calibrations  based on the full data are more accurate than the GaSP calibration using the aggregated data. This is because the true model contains the measurement bias and unknown variance of the noise. The inference based on the full data is thus more precise in this scenario. Furthermore, Example \ref{eg:simulation_stack_bias} indicates that we can accurately estimate the measurement bias and model discrepancy functions  based on multiple sources of data.

\begin{table}[t]
	\caption{Input variables, calibration parameters of the geophysical model, and other model parameters  in calibration.}
	\label{tab:description}
	\centering
	\begin{tabular}{ll} \hline
		Input variables ($\mathbf{x}$)  & Description \\ \hline
		${x}_1$  &  East-west spatial coordinate\\
		${x}_2$   &  North-south spatial coordinate\\ \hline
		Calibration parameters ($\bm{\theta}$) & Description \\ \hline
		${\theta}_1 \in [-2000,3000]$   &  East-west spatial coordinate of chamber centroid ($m$)\\
		${\theta}_2 \in [-2000,5000]$&  North-south spatial coordinate of chamber centroid ($m$) \\
		${\theta}_3 \in [500,6000]$  & Depth of the chamber ($m$) \\
		${\theta}_4 \in [0,0.15]$  & Volume change rate of the reservoir ($m^3/s$)\\
		${\theta}_5 \in [0.25,0.33]$  & Poisson's ratio (host rock property)\\ 
		\hline
		Model parameters  & Description \\
		\hline
		$\bm \mu=(\mu_1,...,\mu_k)$          & Mean parameters \\ 
        $\bm \beta_{1:k}=(\beta_{1,1},\beta_{1,2},...,\beta_{k,1},\beta_{k,2})$     & Inverse range parameters of the measurement bias  \\
		$\bm \eta=(\eta_{1},...,\eta_{k})$ & Nugget parameters of the measurement bias  \\
		$\bm \sigma^2=(\sigma^2_1,...,\sigma^2_k)$  & Scale parameters of the measurement bias \\
		$\bm \beta=(\beta_1,\beta_2)$ & Range parameters of the model discrepancy \\ 
		 $\tau^2$           & Scale parameter of the model discrepancy \\
		\hline

	\end{tabular}
\end{table}
\section{Model calibration by multiple InSAR interferograms}
\label{sec:real}

\begin{figure}[t]
		\begin{tabular}{c}
		\includegraphics[width=.98\textwidth]{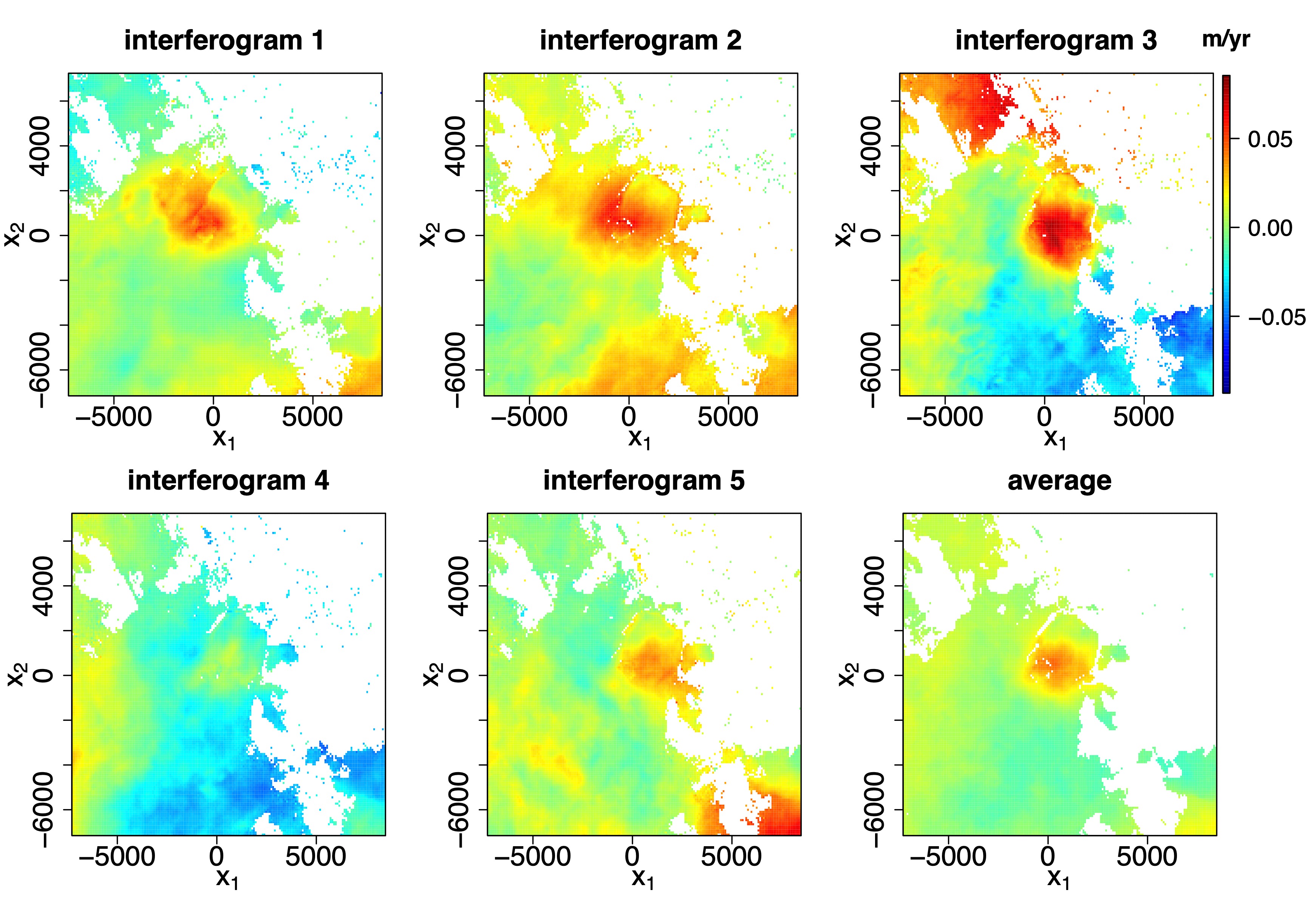}	 \vspace{-.15in}\\

		\end{tabular}
		\caption{Five COSMO-SkyMed satellite interferograms spanning the following time periods: 1) 17 Oct 2011 - 04 May 2012; 2) 21 Oct 2011 - 16 May 2012; 3); 20 Oct 2011 to 15 May 2012; 4) 28 Oct 2011 to 11 May 2012; 5) 12 Oct 2011 - 07 May 2012.  Interferograms 1 and 2  have an ascending-mode look angle, while the rest are descending-mode.
		Horizontal position is in meters relative to a chosen point in K\={\i}lauea Caldera; vertical scale is m/yr. The last figure shows the stack (average) of 6 images. 
		}
		\label{fig:real_images}
	\end{figure}

InSAR data have been widely used at K\={\i}lauea and other volcanoes to estimate the locations and volume changes of magma reservoirs and intrusions \citep[e.g.,][]{Poland2014b, anderson2019magma}. In this section, we study the performance of the aforementioned approaches in calibrating a geophysical model of K\={\i}lauea Volcano using interferograms spanning late-2011 to mid-2012 (Figure \ref{fig:real_images}). During this time, the summit of the volcano inflated due to the storage of magma supplied from the Earth’s mantle \citep{anderson2016bayesian}. Our goal is to use InSAR observations to obtain an improved characterization of the location of the magma reservoir and its volume change, which is important for hazard assessments and for resolving the rate of magma supply to the volcano. Our work extends previous analysis \citep{anderson2016bayesian} by utilization of additional interferograms and consideration of spatially correlated data, noise, and discrepancy functions. 

Consistent with past work, we model the InSAR data by a  geophysical model of volume change of a spherical magma reservoir embedded in an elastic medium \citep{Mogi1958,anderson2016bayesian}, parameterized by the 3D location of its centroid (east distance, north distance, and depth), its volume change, and the Poisson's ratio of the elastic medium.  The input variables, calibration parameters, and other model parameters are listed in Table \ref{tab:description} for calibrating the geophysical model of K\={\i}lauea Volcano using multiple interferograms.


Ground deformation velocities computed from five interferograms captured by the COSMO-SkyMed satellite spanning late 2011 to mid-2012 are shown in Figure \ref{fig:real_images} (in our model, the rate of ground deformation is assumed constant over the complete time range, and the small misalignment between the start and end dates across different interferograms should have only negligible effects). We notice a relatively large measurement bias from the atmospheric error in  interferograms, which makes using multiple interferograms necessary for model calibration.  

We use model (\ref{equ:main_model}), which includes the measurement bias term and the discrepancy function, and we compare the difference between using a GaSP and an S-GaSP model for the discrepancy function.
The range of the calibration parameters using GaSP and S-GaSP calibrations are given in Table \ref{tab:description}.
In both models, 50,000 posterior samples are drawn with the first 10,000 posterior samples used as the burn-in samples. The posterior samples in every 10th step are saved to reduce storage space and autocorrelation in  Markov chains. We present the results based on 400 uniformly sampled pixels here, and the results using quadtree subsampling and stacked interferograms in Section S6 
 in the supplementary materials.  %

 	  \begin{figure}[t]
		\begin{tabular}{c}
		   \includegraphics[width=1\textwidth]{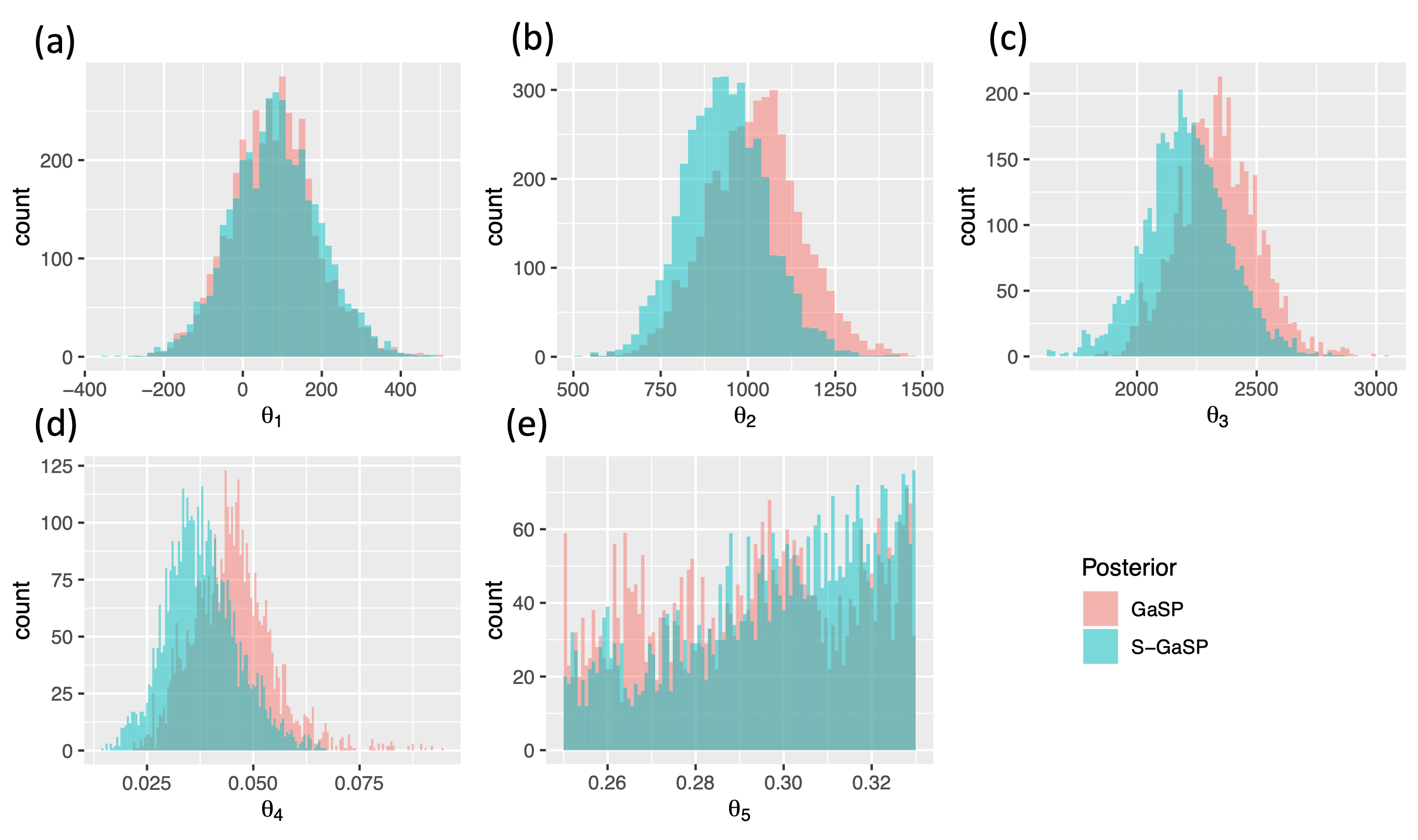}	 \vspace{-.15in}
		\end{tabular}
		\caption{The posterior samples of $\bm \theta$ in the GaSP and S-GaSP calibrations. The range of the  parameter $\theta_5$ (Poisson's ratio) is consistent with many rock types, but the geophysical model is relatively insensitive to this parameter \citep{Mogi1958}.}
		\label{fig:posterior_histogram}
	\end{figure}
Figure \ref{fig:posterior_histogram} graphs the posterior samples of the model calibration parameters. Estimates of the chamber depth ($\theta_3$) and volume change rate ($\theta_4$) are  larger when the discrepancy function is modeled using a GaSP than with an S-GaSP. 
This is because, for a given variance, the GaSP prior places a large probability mass on smooth discrepancy functions. Here the deeper magma chamber and larger volume change rate from the GaSP calibration yield a discrepancy function with relatively smooth ground deformation over a large region. In comparison, the estimated depth and volume change rate by S-GaSP is more consistent with other studies using different sources of data \citep{Poland2014b}, as the S-GaSP prior has more probability mass on smaller $L_2$ loss between reality and the computer model.

	\begin{table}[t]
		\caption{Predictive mean squared error (MSE) in the prediction of the full interferograms using the GaSP and S-GaSP models based on 400 uniformly sampled pixels in each interferogram. MSE$_{f^M}$	is the MSE using the calibrated geophysical model for prediction, and MSE$_{f^M+\delta+\delta_l}$ is the MSE using  the combined calibrated geophysical model, discrepancy function, and measurement bias for prediction. The number is by $10^{-4}$.  Bold font indicates a smaller error. 
		}	
		\label{tab:prediction_real_data}
		\centering
		\begin{tabular}{lcccccc}
			\hline
		MSE$_{f^M}$		& image 1 & image 2  & image 3 & image 4 & image 5\\
			\hline  
				GaSP  & 	 $1.26$  & $1.63$ & $7.80$ & $4.33$ & $1.97$ \\
S-GaSP  & 	 ${\bf 1.21}$  & ${\bf 1.45}$ & ${\bf 7.66}$ & ${\bf 4.05}$ & ${\bf1.76}$  \\
			\hline  
		MSE$_{f^M+\delta+\delta_l}$		& image 1 & image 2  & image 3 & image 4 & image 5  \\
			\hline  
	GaSP  & 	 $ 0.116$  & $ 0.115$ & ${\bf 0.264}$ & $0.134$ & $\bf 0.120$ \\
	S-GaSP  & 	 ${\bf 0.109}$  & ${\bf 0.112}$ & $ 0.267$ & ${\bf 0.131}$ & $0.123$  \\
			\hline
		\end{tabular}
	\end{table}

\begin{figure}[t]
		\begin{tabular}{c}
			\includegraphics[width=.98\textwidth]{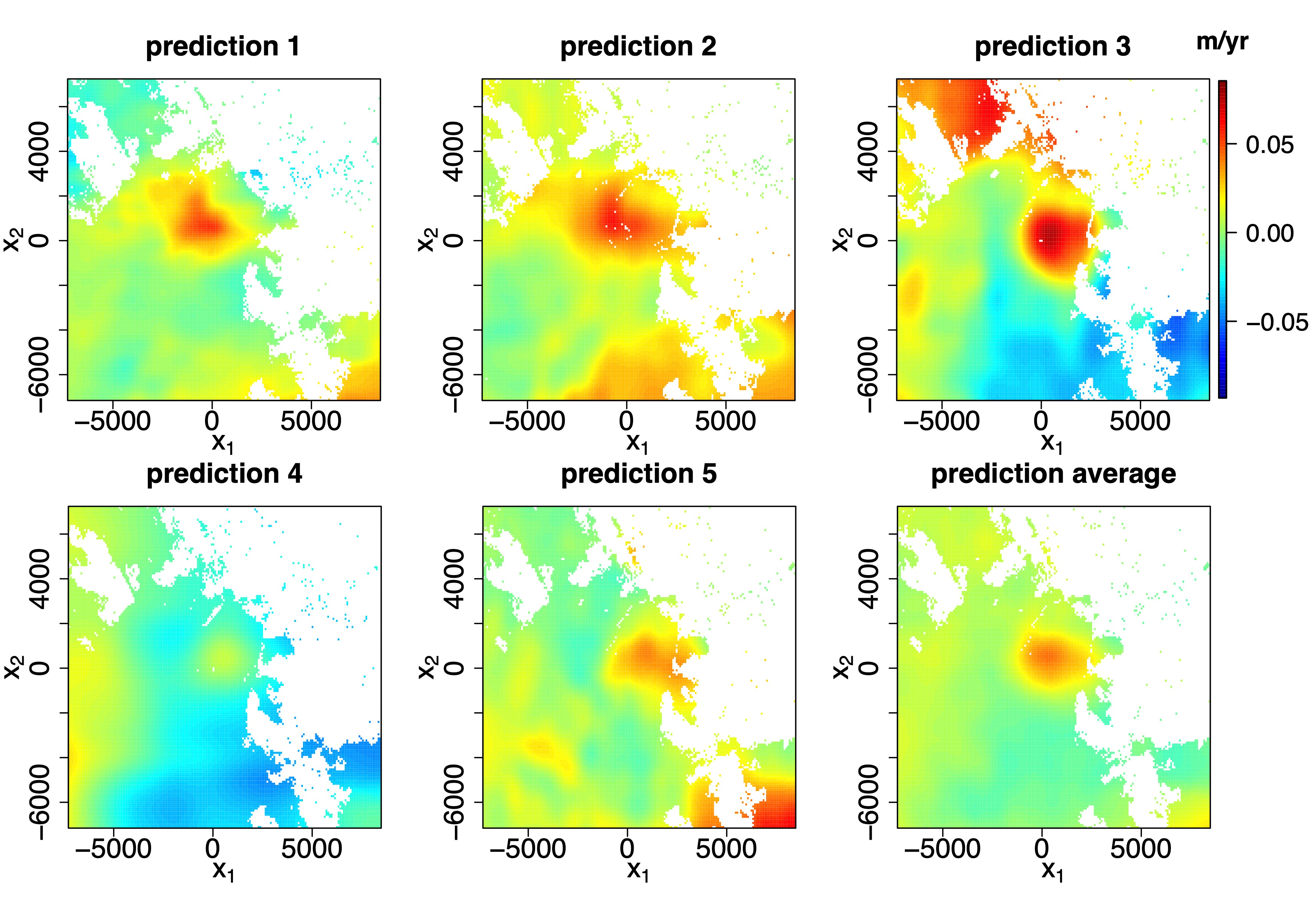}	  \vspace{-.15in}\\
		\end{tabular}
		\caption{Predictive mean of each interferogram and stack image in S-GaSP calibration. }
		\label{fig:SGaSP_prediction}
	\end{figure}

	The MSE of the predictions on each interferogram based on 400 uniformly sampled pixels are given in Table \ref{tab:prediction_real_data}. The mean parameters $\bm \mu$  are treated as a part of the geophysical model for making predictions.
	As shown in Figure \ref{fig:SGaSP_prediction} and Figure S3, the predictive mean using both the calibrated  computer model, discrepancy, and measurement bias terms is accurate in both GaSP and S-GaSP calibration. However, the computer model calibrated by S-GaSP  is more accurate than the GaSP, as shown in Figure S6 
	in the supplementary materials.	In S-GaSP calibration, the prior of the discrepancy has more probability mass near zero, allowing the calibrated geophysical model to explain more variability in the observations than GaSP calibration. 
Consequently, the calibrated geophysical model by S-GaSP is more accurate in prediction  than the GaSP-calibrated model, shown in the first two rows of Table \ref{tab:prediction_real_data}. Furthermore, we also explored other pioneering methods, such as LS and $L_2$ calibration \citep{tuo2015efficient,wong2017frequentist}. However, the convergence of numerical optimization seems to be a challenging issue when there are multiple sources of data. In addition, the measurement bias, which is the focus of this work, was  not considered in previous methods.

	The predictive mean of each interferogram and stacked (averaged) interferogram from the S-GaSP calibration is shown in Figure \ref{fig:SGaSP_prediction}. Predictions are very close to the real interferogram (with around $99\%$ of pixels held out), and performance is better than the  GaSP calibration method. In Section S6 of the supplementary materials, we provide estimated measurement bias and model discrepancy for GaSP and S-GaSP calibrations, and trace plots of all the parameters.

Finally, 
we  compare our results with previous studies of K\={\i}lauea Volcano. The second and third interferograms shown in Figure \ref{fig:real_images} were also used for calibrating the same geophysical model as part of a broader geophysical study \citep{anderson2016bayesian}. However, that work did not consider spatially correlated noise in the data or a discrepancy function.  The same two interferograms were used in \cite{gu2018sgasp} for calibration with a discrepancy function, but the interferogram measurement bias was neglected. Of all the images, the ones used in the previous studies show the largest apparent volcanic ground displacement. As a result, the reservoir volume change rate ($\theta_4$) we estimate here in the S-GaSP calibration using all five images is smaller than in those studies (0.02~m$^3$/s vs. 0.04--0.05~m$^3$/s, respectively). 

The estimated reservoir position depends on the spatial pattern of displacement but not the rate. We estimate a reservoir location $\sim$500~m east and $\sim$800~m north of the reference position (southeast rim of Halema`uma`u Crater) at 1.9 km depth. Many previous studies have examined reservoir locations using a variety of data sets over many years. Despite the relatively low signal-to-noise ratio in the data, our estimated depth is consistent with these studies (e.g., \cite{Poland2014b}). The horizontal position of our most likely reservoir centroid is several hundred meters north of the most commonly accepted location near the east rim of Halema`uma`u Crater. However, it is closer than the position estimated previously in \cite{anderson2016bayesian} and \cite{gu2018sgasp} using two interferograms without modeling the measurement bias; this confirms the importance of addressing the uncertainty in the measurement bias. 
Future studies may combine not only multiple measurements of a single type but also multiple types of data (for instance, GPS or ground tilt), and  
may also utilize more sophisticated geophysical models with larger numbers of calibration parameters. Furthermore, advanced sampling algorithms and design techniques (e.g., \cite{mak2018support}) may be used to represent the posterior distribution.



	\section{Concluding remarks}
	\label{sec:conclusion}
	
	We have introduced a statistical framework to estimate measurement bias, model discrepancy, and calibration parameters using multiple sources of data. In addition, we studied  data reduction by aggregating different sources of data and reducing the number of observations in one source of data.
	We have shown that modeling the full data is more efficient than the aggregated data when either variance of the noise or measurement bias is unknown. We have also shown that certain data reduction approaches, such as the quadtree algorithm, can be very useful for reducing the computational cost of modeling the InSAR interferograms for volcanic hazard quantification. Numerical results based on simulated experiments and real observations validate these findings. 
	
	There are several possible future extensions. First, the computation based on the full data scales linearly with the number of data sources when the inputs (i.e., spatial coordinates in InSAR interferograms) are aligned. When the inputs are misaligned, it will be helpful to design an algorithm for scalable computation. Second, quadtree processing is used widely to downsample satellite interferograms. A theoretical study on how to model quadtree images that properly takes into account the size of the boxes and the measurement bias will be beneficial. It will also be interesting to study  whether quadtree-processed images improve calibration and prediction accuracy compared to alternate designs \citep[e.g.,][]{fukushima2005finding}.  Lastly, for volcanological applications,
	more work is required to fuse diverse data types, such as gas emissions, GPS data, and InSAR data, with geophysical models for Bayesian inversion. 
    
    \section*{Supplementary materials}
    The supplementary materials contain 6 sections. Section S1 and S2 give the proof of Lemma 1 and derivations of the marginal likelihood and predictive distribution of the calibration model, respectively. An example to illustrate the inconsistent maximum likelihood estimation of the GaSP calibration model is discussed Section S3. We introduce the discretization of S-GaSP calibration model in Section S4, and provide additional numerical results for simulation studies and real data analysis in Section S5 and Section S6, respectively.

\section*{Acknowledgements}
This research is supported by the National Science
Foundation under Award No. DMS-2053423. The authors thank the editor, AE, Chuck Wicks, and two referees for their comments that substantially improved the article. Any use of trade, firm, or product names is for descriptive purposes only and does not imply endorsement by the U.S. Government.  

	\bibliographystyle{apalike}
	\bibliography{References_2018}

\begin{thebibliography}{}

\bibitem[Anderson and Segall, 2011]{Anderson2011}
Anderson, K. and Segall, P. (2011).
\newblock Physics-based models of ground deformation and extrusion rate at
  effusively erupting volcanoes.
\newblock {\em Journal of Geophysical Research: Solid Earth}, 116(B7):B07204.

\bibitem[Anderson et~al., 2019]{anderson2019magma}
Anderson, K.~R., Johanson, I.~A., Patrick, M.~R., Gu, M., Segall, P., Poland,
  M.~P., Montgomery-Brown, E.~K., and Miklius, A. (2019).
\newblock Magma reservoir failure and the onset of caldera collapse at
  {K}{\=\i}lauea volcano in 2018.
\newblock {\em Science}, 366(6470):eaaz1822.

\bibitem[Anderson and Poland, 2016]{anderson2016bayesian}
Anderson, K.~R. and Poland, M.~P. (2016).
\newblock Bayesian estimation of magma supply, storage, and eruption rates
  using a multiphysical volcano model: K{\=\i}lauea volcano, 2000--2012.
\newblock {\em Earth and Planetary Science Letters}, 447:161--171.

\bibitem[Arendt et~al., 2012a]{arendt2012quantification}
Arendt, P.~D., Apley, D.~W., and Chen, W. (2012a).
\newblock Quantification of model uncertainty: Calibration, model discrepancy,
  and identifiability.
\newblock {\em Journal of Mechanical Design}, 134(10):100908.

\bibitem[Arendt et~al., 2012b]{arendt2012improving}
Arendt, P.~D., Apley, D.~W., Chen, W., Lamb, D., and Gorsich, D. (2012b).
\newblock Improving identifiability in model calibration using multiple
  responses.
\newblock {\em Journal of Mechanical Design}, 134(10):100909.

\bibitem[Bayarri et~al., 2007]{bayarri2007framework}
Bayarri, M.~J., Berger, J.~O., Paulo, R., Sacks, J., Cafeo, J.~A., Cavendish,
  J., Lin, C.-H., and Tu, J. (2007).
\newblock A framework for validation of computer models.
\newblock {\em Technometrics}, 49(2):138--154.

\bibitem[B{\"{u}}rgmann et~al., 2000]{Burgmann2000}
B{\"{u}}rgmann, R., Rosen, P.~A., and Fielding, E.~J. (2000).
\newblock {Synthetic Aperture Radar Interferometry to Measure Earth's Surface
  Topography and Its Deformation}.
\newblock {\em Annual Review of Earth and Planetary Sciences}, 28(1):169--209.

\bibitem[Casella and Berger, 2002]{casella2021statistical}
Casella, G. and Berger, R.~L. (2002).
\newblock {\em Statistical Inference, 2nd Edition}.
\newblock Cengage Learning.

\bibitem[Chen and Zebker, 2001]{Chen2001}
Chen, C.~W. and Zebker, H.~A. (2001).
\newblock {Two-dimensional phase unwrapping with use of statistical models for
  cost functions in nonlinear optimization}.
\newblock {\em Journal of the Optical Society of America A}, 18(2):338.

\bibitem[Fukushima et~al., 2005]{fukushima2005finding}
Fukushima, Y., Cayol, V., and Durand, P. (2005).
\newblock Finding realistic dike models from interferometric synthetic aperture
  radar data: The {F}ebruary 2000 eruption at {P}iton de la {F}ournaise.
\newblock {\em Journal of Geophysical Research: Solid Earth}, 110(B3):B03206.

\bibitem[Gong et~al., 2016]{Gong2016}
Gong, W., Lu, Z., and Meyer, F. (2016).
\newblock {Uncertainties in estimating magma source parameters from InSAR
  observation}.
\newblock In Riley, K., Webley, P., and Thompson, M., editors, {\em Natural
  Hazard Uncertainty Assessment: Modeling and Decision Support, Geophysical
  Monograph 223}, chapter~7, pages 89--104. John Wiley {\&} Sons, Inc.

\bibitem[Gu, 2018]{gu2018jointly}
Gu, M. (2018).
\newblock Jointly robust prior for {G}aussian stochastic process in emulation,
  calibration and variable selection.
\newblock {\em Bayesian Analysis}, 14(1).

\bibitem[Gu, 2022]{gu2022robustcalibration}
Gu, M. (2022).
\newblock Robustcalibration: Robust calibration of computer models in {R}.
\newblock {\em arXiv preprint arXiv:2201.01476}.

\bibitem[Gu and Wang, 2018]{gu2018sgasp}
Gu, M. and Wang, L. (2018).
\newblock Scaled {G}aussian stochastic process for computer model calibration
  and prediction.
\newblock {\em SIAM/ASA Journal on Uncertainty Quantification},
  6(4):1555--1583.

\bibitem[Gu et~al., 2022]{gu2018theoretical}
Gu, M., Xie, F., and Wang, L. (2022).
\newblock A theoretical framework of the scaled {G}aussian stochastic process
  in prediction and calibration.
\newblock {\em SIAM/ASA Journal on Uncertainty Quantification},
  10(4):1435--1460.

\bibitem[Handcock and Stein, 1993]{handcock1993bayesian}
Handcock, M.~S. and Stein, M.~L. (1993).
\newblock A {B}ayesian analysis of {K}riging.
\newblock {\em Technometrics}, 35(4):403--410.

\bibitem[Higdon et~al., 2008]{higdon2008computer}
Higdon, D., Gattiker, J., Williams, B., and Rightley, M. (2008).
\newblock Computer model calibration using high-dimensional output.
\newblock {\em Journal of the American Statistical Association},
  103(482):570--583.

\bibitem[Hodges and Reich, 2010]{hodges2010adding}
Hodges, J.~S. and Reich, B.~J. (2010).
\newblock Adding spatially-correlated errors can mess up the fixed effect you
  love.
\newblock {\em The American Statistician}, 64(4):325--334.

\bibitem[Hooper et~al., 2007]{Hooper2007}
Hooper, A., Segall, P., and Zebker, H. (2007).
\newblock {Persistent Scatterer InSAR for Crustal Deformation Analysis, with
  Application to Volcan Alcedo, Galapagos}.
\newblock {\em Journal of Geophysical Research}, 112:1--19.

\bibitem[Jonsson et~al., 2002]{Jonsson2002}
Jonsson, S., Zebker, H., Segall, P., and Amelung, F. (2002).
\newblock {Fault Slip Distribution of the 1999 Mw 7.1 Hector Mine, California,
  Earthquake, Estimated from Satellite Radar and GPS Measurements}.
\newblock {\em Bulletin of the Seismological Society of America},
  92(4):1377--1389.

\bibitem[Joseph and Mak, 2021]{joseph2021supervised}
Joseph, V.~R. and Mak, S. (2021).
\newblock Supervised compression of big data.
\newblock {\em Statistical Analysis and Data Mining: The ASA Data Science
  Journal}, 14(3):217--229.

\bibitem[Kennedy and O'Hagan, 2001]{kennedy2001bayesian}
Kennedy, M.~C. and O'Hagan, A. (2001).
\newblock Bayesian calibration of computer models.
\newblock {\em Journal of the Royal Statistical Society: Series B (Statistical
  Methodology)}, 63(3):425--464.

\bibitem[Knospe and Jonsson, 2010]{Knospe2010}
Knospe, S. and Jonsson, S. (2010).
\newblock Covariance estimation for dinsar surface deformation measurements in
  the presence of anisotropic atmospheric noise.
\newblock {\em IEEE Transactions on Geoscience and Remote Sensing},
  48(4):2057--2065.

\bibitem[Lohman and Simons, 2005]{Lohman2005}
Lohman, R.~B. and Simons, M. (2005).
\newblock Some thoughts on the use of insar data to constrain models of surface
  deformation: Noise structure and data downsampling.
\newblock {\em Geochemistry, Geophysics, Geosystems}, 6(1):Q01007.

\bibitem[Mak and Joseph, 2018]{mak2018support}
Mak, S. and Joseph, V.~R. (2018).
\newblock Support points.
\newblock {\em Annals of Statistics}, 46(6A):2562--2592.

\bibitem[Massonnet et~al., 1995]{Massonnet1995}
Massonnet, D., Briole, P., and Arnaud, A. (1995).
\newblock {Deflation of Mount Etna monitored by spaceborne radar
  interferometry}.
\newblock {\em Nature}, 375(6532):567--570.

\bibitem[Mogi, 1958]{Mogi1958}
Mogi, K. (1958).
\newblock {Relations between the eruptions of various volcanoes and the
  deformations of the ground surfaces around them}.
\newblock {\em Bulletin of the Earthquake Research Institute, University of
  Tokyo}, 36:99--134.

\bibitem[Montgomery-Brown et~al., 2015]{Montgomery-Brown2015a}
Montgomery-Brown, E.~K., Wicks, C.~W., Cervelli, P.~F., Langbein, J.~O., Svarc,
  J.~L., Shelly, D.~R., Hill, D.~P., and Lisowski, M. (2015).
\newblock {Renewed inflation of Long Valley Caldera, California (2011 to
  2014)}.
\newblock {\em Geophysical Research Letters}, 42(13):5250--5257.

\bibitem[Neal et~al., 2019]{neal20192018}
Neal, C.~A., Brantley, S., Antolik, L., Babb, J., Burgess, M., Calles, K.,
  Cappos, M., Chang, J., Conway, S., Desmither, L., et~al. (2019).
\newblock The 2018 rift eruption and summit collapse of {K}{\=\i}lauea volcano.
\newblock {\em Science}, 363(6425):367--374.

\bibitem[Pinel et~al., 2014]{Pinel2014}
Pinel, V., Poland, M., and Hooper, A. (2014).
\newblock {Volcanology: Lessons learned from Synthetic Aperture Radar imagery}.
\newblock {\em Journal of Volcanology and Geothermal Research}, 289:81--113.

\bibitem[Plumlee, 2017]{plumlee2016bayesian}
Plumlee, M. (2017).
\newblock Bayesian calibration of inexact computer models.
\newblock {\em Journal of the American Statistical Association},
  112(519):1274--1285.

\bibitem[Poland et~al., 2014]{Poland2014b}
Poland, M.~P., Miklius, A., and Montgomery-Brown, E.~K. (2014).
\newblock {Magma Supply, Storage, and Transport at Shield-Stage Hawaiian
  Volcanoes}.
\newblock In {\em Characteristics of Hawaiian Volcanoes}, chapter~5, pages
  179--234. U.S. Geological Survey Professional Paper 1801.

\bibitem[Pritchard et~al., 2002]{Pritchard2002}
Pritchard, M., Simons, M., Rosen, P., Hensley, S., and Webb, F. (2002).
\newblock Co-seismic slip from the 1995 {J}uly 30 {M}w = 8.1 {A}ntofagasta,
  {C}hile, earthquake as constrained by {InSAR} and {GPS} observations.
\newblock {\em Geophysical Journal International}, 150(2):362--376.

\bibitem[Rasmussen, 2006]{rasmussen2006gaussian}
Rasmussen, C.~E. (2006).
\newblock {\em Gaussian processes for machine learning}.
\newblock MIT Press.

\bibitem[Reich et~al., 2006]{reich2006effects}
Reich, B.~J., Hodges, J.~S., and Zadnik, V. (2006).
\newblock Effects of residual smoothing on the posterior of the fixed effects
  in disease-mapping models.
\newblock {\em Biometrics}, 62(4):1197--1206.

\bibitem[Simons et~al., 2002]{Simons2002b}
Simons, M., Fialko, Y., and Rivera, L. (2002).
\newblock {Coseismic deformation from the 1999 Mw 7.1 Hector Mine, California,
  earthquake as inferred from InSAR and GPS observations}.
\newblock {\em Bulletin of the Seismological Society of America},
  92(4):1390--1402.

\bibitem[Simons and Rosen, 2007]{Simons2007}
Simons, M. and Rosen, P.~A. (2007).
\newblock {Interferometric synthetic aperture radar geodesy}.
\newblock In {\em Treatise on Geophysics}, volume~3, pages 391--446. Elsevier
  Press.

\bibitem[Tuo and Wu, 2015]{tuo2015efficient}
Tuo, R. and Wu, C.~J. (2015).
\newblock Efficient calibration for imperfect computer models.
\newblock {\em The Annals of Statistics}, 43(6):2331--2352.

\bibitem[Tuo and Wu, 2016]{tuo2016theoretical}
Tuo, R. and Wu, C.~J. (2016).
\newblock A theoretical framework for calibration in computer models:
  parametrization, estimation and convergence properties.
\newblock {\em SIAM/ASA Journal on Uncertainty Quantification}, 4(1):767--795.

\bibitem[Wang et~al., 2014]{Wang2014}
Wang, C., Ding, X., Li, Q., and Jiang, M. (2014).
\newblock {Equation-based InSAR data quadtree downsampling for earthquake slip
  distribution inversion}.
\newblock {\em IEEE Geoscience and Remote Sensing Letters}, 11(12):2060--2064.

\bibitem[Wang et~al., 2020]{wang2020prediction}
Wang, W., Tuo, R., and Jeff~Wu, C. (2020).
\newblock On prediction properties of kriging: Uniform error bounds and
  robustness.
\newblock {\em Journal of the American Statistical Association},
  115(530):920--930.

\bibitem[Wong et~al., 2017]{wong2017frequentist}
Wong, R.~K., Storlie, C.~B., and Lee, T. (2017).
\newblock A frequentist approach to computer model calibration.
\newblock {\em Journal of the Royal Statistical Society: Series B (Statistical
  Methodology)}, 79:635--648.

\bibitem[Zebker et~al., 1997]{Zebker1997}
Zebker, H.~A., Rosen, P.~A., and Hensley, S. (1997).
\newblock {Atmospheric effects in interferometric synthetic aperture radar
  surface deformation and topographic maps}.
\newblock {\em Journal of Geophysical Research: Solid Earth},
  102(B4):7547--7563.

\bibitem[Zhang, 2004]{zhang2004inconsistent}
Zhang, H. (2004).
\newblock Inconsistent estimation and asymptotically equal interpolations in
  model-based geostatistics.
\newblock {\em Journal of the American Statistical Association},
  99(465):250--261.

\end{thebibliography}
\end{document}